\documentclass[aps,floats]{revtex4}
\usepackage{amsmath,amssymb}
\usepackage{graphicx,epsfig}
\usepackage[greek,english]{babel}
\usepackage{bbold}
\begin{document}
\bibliographystyle {plain}

\pdfoutput=1
\def\oppropto{\mathop{\propto}} 
\def\opsimeq{\mathop{\simeq}}
\def\opoverderline{\mathop{\overline}}
\def\operarrow{\mathop{\longrightarrow}}
\def\opsim{\mathop{\sim}}

\def\fig#1#2{\includegraphics[height=#1]{#2}}
\def\figx#1#2{\includegraphics[width=#1]{#2}}

%\newcommand{\fig}[2]{\epsfxsize=#1\epsfbox{#2}} \reversemarginpar 

%%%%%%%%%%%%%%%%%%%%%%%%%%%%%%%%%%%%%%%%%%%%%%%%%%%%%%%%%%%%%%%%%%%%%%%%%%%%
\title{ On various levels of deterministic toy models for the Richardson cascade in turbulence  } 

%%%%%%%%%%%%%%%%%%%%%%%%%%%%%%%%%%%%%%%%%%%%%%%%%%%%%%%%%%%%%%%%%%%%%%%%%%%%

\author{ C\'ecile Monthus }
 \affiliation{Institut de Physique Th\'{e}orique, 
Universit\'e Paris Saclay, CNRS, CEA,
91191 Gif-sur-Yvette, France}

\begin{abstract}
The Desnyanski-Novikov shell model  is a deterministic dynamical model for scalar velocities $v_t(n)$
defined on the one-dimensional-lattice $n=0,1,2,..$ labelling the length-scales $l_n=l_0 2^{-n}$,
in order to describe the cascade of energy from the biggest scale where it is injected by some external forcing
towards the smaller scales where it is dissipated by viscosity. We describe the generalization of this model
in two directions :  (i) the one-dimensional-lattice $n=0,1,2,..$ labelling the length-scales $l_n=l_0 2^{-n}$
is replaced by a scale-spatial tree structure of nested cells in order to allow spatial heterogeneities between different coherent structures
that are localized in different regions of the whole volume ; (ii) the scalar velocities $v_t(n)$ are replaced by 3D-vector velocities 
in order to take into account the vorticity in the dynamical equations and to include vortex-stretching effects.

\end{abstract}

\maketitle

%%%%%%%%%%%%%%%%%%%%%%%%%%%%%%%%%%%%%%%%%%%%%%%%%%%%%%%%%%%%%%%

\section{ Introduction }

The study of cascades of energy and other conserved observables in various turbulent flows
has remained a very active field over the years
(see the recent review \cite{review_cascade} and references therein). 
Among the various approaches to isotropic fully developed turbulence in dimension $D=3$ (see the book \cite{frisch} and references therein),
shell models \cite{book_shell} are deterministic toy models that try to keep some Navier-Stokes flavor.
The general idea is to replace the Navier-Stokes partial differential equation for the $3D$ velocity field $\vec v (\vec r)$
by a finite number of coupled nonlinear ordinary differential equations in continuous time for the velocities of some selected modes.
The simplest example is the Desnyanski-Novikov shell model  
\cite{obu,DN,gg,dombre}
where each length-scale $l_n=l_0 2^{-n}$ labelled by the integer $n=0,1,2,..$
is characterized by a single real velocity $v_t(n)$
and where the interactions between velocities are limited to nearest neighbors (see more details in section \ref{sec_DN}).
Another famous example is the so-called GOY model reviewed in \cite{goy_review}
where the velocity $v_t(n)$ is instead  a complex variable,
and where the interactions between velocities include next-nearest-neighbors.

This GOY model defined on the one-dimensional scale lattice $n$
has been then extended to a scale-spatial tree structure \cite{benzi_1p1,benzi_inter,benzi_ultra},
in order to take into account the real-space physical picture of the Richardson cascade with localized eddies of various sizes 
and to allow for spatial heterogeneities : 
in the simplest setting, each eddy interacts only with its ancestors which spatially contain it
and with its children spatially contained in it (more complex choices have been also studied in \cite{benzi_1p1,benzi_inter,benzi_ultra}).
At the heuristic level, the velocities on this scale-spatial tree structure
should be considered as the coefficients of an orthonormal wavelet expansion of the real velocity field 
of fluid mechanics \cite{benzi_1p1,benzi_inter,benzi_ultra}),
even if one does not even try to write the explicit wavelet basis that would have these velocities as coefficients.
Note that the idea that wavelets are the appropriate language to analyse turbulence
had appeared previously \cite{argoul,muzy,farge,arneodo} independently of shell models.

In the present paper, our goal is to apply this scale-spatial tree construction to the Desnyanski-Novikov shell model  
 and to consider also the generalization to velocity vectors in $D=3$
in order to take into account the vorticity in the dynamical equations and to include vortex-stretching effects.
The paper is organized as follows.
In section \ref{sec_DN}, we revisit the Desnyanski-Novikov shell model for scalar velocities $v_t(n)$ 
one the 1-dimensional scale-lattice $n$, in order to stress that the Liouville phase-space-volume-conservation 
fixes the relation between the two parameters $(\alpha,\beta)$ that are usually considered as independent.
In section \ref{sec_tree}, we describe the generalization where the one the 1-dimensional scale-lattice $n$
is replaced by the scale-spatial tree structure.
In section \ref{sec_3D}, we describe the generalization where the scalar velocities $v_t(n)$ 
are replaced by 3D vectors $\vec v_t(n)$.
In section \ref{sec_3Dtree}, the two generalizations are put together to obtain
the dynamical model for 3D velocities on the scale-spatial tree structure.
Our conclusions are summarized in section \ref{sec_conclusion}.
In  Appendix \ref{app_navier}, we recall the properties of the Navier-Stokes dynamics
in order to stress the correspondences with the shell models discussed in the text.

\section{ Revisiting the Desnyanski-Novikov shell model  }

\label{sec_DN}

In this section, we revisit the Desnyanski-Novikov shell model that will be the starting point for the generalizations
introduced in the next sections.

\subsection{ Dynamics for scalar velocities $v_t(n)$ on the 1-dimensional scale-lattice $n$ }

Each generation $n=0,1,2,..$ corresponding to the length-scale
\begin{eqnarray}
l_n=l_0 2^{-n}
\label{ln}
\end{eqnarray}
is characterized by a real velocity $v_t(n)$ evolving with the dynamics for $n \geq 1$
\begin{eqnarray}
 \frac{d v_t(n)  }{dt}  && =Q_{\alpha,\beta} [v_t(n-1); v_t(n);v_t(n+1) ]  -  \frac{\nu D }{l_n^2} v_t(n) 
\label{vshell1d}
\end{eqnarray}
where $\nu$ is the viscosity that introduces dissipation, 
while the term $Q_{\alpha,\beta}[v_t(n-1); v_t(n);v_t(n+1) ]  $ 
represents the non-linear quadratic interactions between the nearest-neighbors velocities
in the one-dimensional scale-spatial of generations
\begin{eqnarray}
Q_{\alpha,\beta}[v_t(n-1); v_t(n);v_t(n+1) ]  && \equiv \alpha \left( \frac{v_t^2(n-1) }{l_n} - \frac{v_t(n) v_t(n+1) }{l_{n+1}}   \right) 
+  \beta \left(\frac{v_t(n-1) v_t(n) }{l_{n-1}} - \frac{v_t^2(n+1) }{l_n}   \right) 
\label{qab}
\end{eqnarray}
and is meant to mimic as much as possible the conservative part of the Navier-Stokes Equation (recalled in Eqs \ref{navier} \ref{navierlamb} of the Appendix).
The dynamics for the generation $n=0$ has no ancestor at $n=-1$
but includes instead some external forcing $f[v_t(0)]$
\begin{eqnarray}
 \frac{d v_t(0)  }{dt}  && =Q_{\alpha,\beta} [0; v_t(0);v_t(1) ]  -  \frac{\nu D }{l_0^2} v_t(0)  +f[v_t(0)]
\nonumber \\
&& =- \alpha \frac{v_t(0) v_t(1) }{l_{1}}   
-  \beta  \frac{v_t^2(1) }{l_0}  
 -  \frac{\nu D }{l_0^2} v_t(0)  +f[v_t(0)]
\label{vshell1d0}
\end{eqnarray}

\subsection{ Dynamics for the kinetic energies  }

The corresponding kinetic energies
\begin{eqnarray}
 e_t(n) \equiv \frac{ v_t^2(n)  }{2}  
\label{etndef}
\end{eqnarray}
then evolves for $n \geq 1$ according to
\begin{eqnarray}
 \frac{d e_t(n) }{dt}   && = v_t(n)\frac{d v_t(n)  }{dt}  
 =   j_t(n-1 \to n)  - j_t(n \to n+1)    -  \frac{ 2\nu  }{l_n^2} e_t(n) 
\label{eshell1d}
\end{eqnarray}
where the energy current from generation $(n-1)$ to generation $n$
\begin{eqnarray}
j_t(n-1 \to n)  \equiv \alpha  v_t^2(n-1) \frac{ v_t(n) }{l_n}  +  \beta \frac{v_t(n-1) }{l_{n-1}} v^2_t(n) 
\label{jeshell1d}
\end{eqnarray}
has a very straightforward physical interpretation,
in terms of the two energies $v_t^2(n- 1)$ and $v^2_t(n)  $ and the two advection factors $\frac{ v_t(n) }{l_n} $ and $\frac{v_t(n-1) }{l_{n-1}}  $.
Note that in the GOY shell model (see the review \cite{goy_review} and references therein),
the energy current is somewhat less intuitive since it involves the products of velocities over three consecutive shells.
In Eq \ref{jeshell1d}, the signs of the two contributions to this current depend on the signs of the velocities :
the $\alpha$ contribution corresponds to a direct cascade towards smaller scale for positive $(\alpha v_t(n))>0$
and to an inverse cascade towards bigger scales for negative $(\alpha v_t(n))<0$.
Similarly, the $\beta$ contribution corresponds to a direct cascade towards smaller scale for positive $(\beta v_t(n-1))>0$
and to an inverse cascade towards bigger scales for negative $(\beta v_t(n-1))>0$.

For $n=0$, Eq. \ref{vshell1d0} leads to
\begin{eqnarray}
 \frac{d e_t(0)  }{dt}  && =- j_t(0 \to 1)
 -  \frac{\nu D }{l_0^2} v^2_t(0)  + v_t(0)f[v_t(0)]
\label{eshell1d0}
\end{eqnarray}

As a consequence, the dynamics for the total energy
\begin{eqnarray}
e^{tot}_t && \equiv \sum_{n=0}^{N} e_t(n) 
\label{e1d}
\end{eqnarray}
contains only the dissipation over all generations and the power injected by the external forcing on generation $n=0$
\begin{eqnarray}
\frac{d e^{tot}_t}{dt}  && = - 2 \nu D \sum_{n=0}^{+\infty} \frac{ e_t(n)  }{l_n^2}  +v_t(0)f[v_t(0)]
\label{etot1d}
\end{eqnarray}

In the absence of external forcing $f[v_t(0)]=0$ and viscosity $\nu=0$,
the total energy $e^{tot}_t$ 
 is conserved for any values of the parameters $(\alpha,\beta)$, 
that are thus usually considered as two independent free parameters of the model in the literature \cite{frisch,dombre}.
However, as explained in the following section, the supplementary constraint $\alpha=\beta$ appears necessary
if one takes into account the dynamical properties in phase space.

\subsection{ Dynamics for the phase-space-volume 
 in the absence of dissipation ($\nu=0$)
}

In the Hamiltonian formulation of classical mechanics, the Liouville theorem 
concerning the phase-space-volume conservation plays an essential role.
Since the non-dissipative part of the Navier-Stokes equation corresponds to hydrodynamic continuum version of classical mechanics,
it should respect the phase-space-volume conservation, even if it is not very often stressed and if it is not technically obvious as a consequence of the
Eulerian formulation in terms of the velocity field. 
Nevertheless, in numerical space-time re-discretizations of the non-dissipative part of the Navier-Stokes equation,
it is known that it is essential to respect the phase-space-volume conservation \cite{egger,sommer} :
 indeed, if the discretization introduces some spurious phase-space-contraction, 
the dynamics will artificially converge towards a lower-dimensional attractor, while if the discretization introduces some spurious 
phase-space-dilatation, the dynamics will artificially increase the chaotic properties via bigger Lyapunov exponents \cite{sommer}.

In the present shell model, the phase-space of the velocities $(v_t(0),v_t(1),...)$
has for elementary volume 
\begin{eqnarray}
d \tau_t \equiv \prod_{n=0}^{+\infty} dv_t(n)
\label{volumephasespace}
\end{eqnarray}

In the absence of dissipation $\nu=0$,
the phase-space local dilation associated to the velocity $v_t(n)$ evolving with the Dynamics of Eq. \ref{vshell1d} for $n \geq 1$ reads
\begin{eqnarray}
\frac{ \partial \left( \frac{d v_t(n)  }{dt}  \right)  }{\partial v_t(n) } 
&& = - \alpha  \frac{ v_t(n+1) }{l_{n+1}}  
+  \beta \frac{v_t(n-1)  }{l_{n-1}} 
\label{vshell1dila}
\end{eqnarray}
while for $n=0$, Eq \ref{vshell1d0} yields
\begin{eqnarray}
\frac{ \partial \left( \frac{d v_t(0)  }{dt}  \right)  }{\partial v_t(0) } 
&& =- \alpha \frac{ v_t(1) }{l_{1}}      +f'[v_t(0)]
\label{vshell1d0dila}
\end{eqnarray}

As a consequence, the total phase-space dilatation reads
\begin{eqnarray}
 \sum_{n=0}^{+\infty}  \frac{ \partial \left( \frac{d v_t(n)  }{dt}  \right)  }{\partial v_t(n) } 
&&
 = f'[v_t(0)] + \beta \frac{v_t(0)  }{l_{0}}  
+  (\beta- \alpha ) \sum_{n=1}^{\infty}  \frac{ v_t(n) }{l_{n}}  
\label{divjphasespace}
\end{eqnarray}

Our conclusion is thus that one should impose the supplementary constraint
\begin{eqnarray}
\beta=\alpha
\label{betachoice}
\end{eqnarray}
in order to satisfy the phase-space-volume-conservation for the conservative part of the shell dynamics for generations $n \geq 1$.
Note that in the GOY shell model (see the review \cite{goy_review} and references therein),
the Liouville phase-space-volume-conservation is instead satisfied locally since $\frac{d v_t(n)  }{dt}  $
does not even depend at all on $v_t(n) $.

In Eq. \ref{qab}, the
 choice $\beta=\alpha$ actually allows to regroup the terms differently as
\begin{eqnarray}
Q_{\alpha,\alpha}[v_t(n-1); v_t(n);v_t(n+1) ]  && 
= \alpha \left(\frac{v_t(n-1) }{l_{n-1}}  - \frac{v_t(n+1) }{l_{n+1}}   \right) v_t(n) 
+  \alpha \left(\frac{v_t^2(n-1) }{l_n} - \frac{v_t^2(n+1) }{l_n}   \right) 
\label{qaa}
\end{eqnarray}
The first term can be interpreted as the advection of $v_t(n) $ by its two neighboring shells $(n \pm 1)$,
while the second term involves the difference of kinetic energies of the two neighboring shells $(n \pm 1)$.
This form thus mimics more directly the familiar terms present in the Navier-Stokes equation,
either with the advection term (Eq. \ref{navier}) or in the alternative form containing the gradient of kinetic energy (Eq. \ref{navierlamb}).

\subsection{ Regime of fully developed turbulence }

The characteristic advection time scale $\tau^{adv}_t(n)$ associated to the velocity $ v_t (n) $ on scale $l_n$ reads
\begin{eqnarray}
\tau^{adv}_t(n) \equiv \frac{l_n} { \vert v_t(n) \vert}
\label{tauadv}
\end{eqnarray}
The comparison with the dissipation time scale induced by the viscosity $\nu$ on scale $l_n$
\begin{eqnarray}
\tau^{diss}(n)= \frac{l_n^2}{\nu} 
\label{taudiss}
\end{eqnarray}
leads to the introduction of the Reynolds numbers
\begin{eqnarray}
R_t(n) \equiv \frac{l_n v_t(n) } {\nu} = \frac{\tau^{diss}(n) }{\tau^{adv}_t(n) } 
\label{reynoldscell}
\end{eqnarray}
that measure the relative importance of advection with respect to dissipation at generation $n$.
The standard Reynolds number is associated to the integral scale $l_0$ 
and is assumed to be extremely large in fully developed turbulence
\begin{eqnarray}
R_t(0) \equiv \frac{l_0 v_t(0) } {\nu} \gg 1
\label{reynolds0}
\end{eqnarray}
As a consequence, the dissipation will remain negligible with respect to the advection over a very large number of generations $0 \leq n \ll N$.
The generation $N$ corresponds to
the Kolmogorov dissipation scale $\eta=l_N=l_0 2^{-N}$  where the Reynolds number of Eq. \ref{reynoldscell} becomes of order unity
\begin{eqnarray}
R_t(N) \equiv \frac{l_N v_t(N) } {\nu} \simeq 1
\label{reynolds1}
\end{eqnarray}

In the inertial range corresponding to generations $1 \leq n \ll N$,
the energy is neither created nor destroyed, but only transported between generations via the energy current of Eq. \ref{jeshell1d}.
In one looks for some time-independent solution, Eq. \ref{eshell1d} yields that
the stationary energy current $  j(n-1 \to n)$ of Eq. \ref{jeshell1d} should be independent of $n$
and take some value traditionally denoted by $\epsilon$ in turbulence (even if it is of course not meant to be small)
\begin{eqnarray}
\epsilon = j(n-1 \to n)  = \alpha  v^2(n-1) \frac{ v(n) }{l_n}  +  \beta \frac{v(n-1) }{l_{n-1}} v^2(n) 
\label{jeshell1deps}
\end{eqnarray}
For $n=0$, the stationary state of Eq \ref{eshell1d0} then yields that the external forcing should be chosen to be
\begin{eqnarray}
f[v(0)] && =  \frac{\epsilon}{v(0) }
\label{f0}
\end{eqnarray}
in order to ensure that the energy injected per unit time and per unit volume is indeed $\epsilon$.
Eq. \ref{jeshell1deps} yields that
the stationary velocity follows the following power-law with respect to the length $l_n=l_0 2^{-n}$
\begin{eqnarray}
v(n) = v(0) \left( \frac{l_n}{l_0} \right)^h = v(0) 2^{- n h}
\label{vn41}
\end{eqnarray}
with the famous 'K41' (Kolmogorov 1941) scaling exponent 
\begin{eqnarray}
h= \frac{1}{3}
\label{hK41}
\end{eqnarray}
where 
\begin{eqnarray}
\epsilon && = \frac{v^3(0) }{l_0 } \left[   \alpha  2^{\frac{2}{3}}  + \beta 2^{- \frac{2}{3}}  \right]
\label{clhomo}
\end{eqnarray}
determines the velocity $v(0)$ at generation $n=0$ in terms of $\epsilon$.

To analyze the linear stability around this steady-state $v(n)$, 
it is convenient to consider that the dynamics in the inertial range $1 \leq n \leq M-1 \ll N$ is described by Eq. \ref{eshell1d}
\begin{eqnarray}
 \frac{d e_t(n) }{dt}   &&  =   j_t(n-1 \to n)  - j_t(n \to n+1)  
\label{eshell1di}
\end{eqnarray}
with the two boundary equations at $n=0$ and $n=M$
\begin{eqnarray}
 \frac{d e_t(0) }{dt}   &&  =   \epsilon  - j_t(0 \to 1)  
\nonumber \\
 \frac{d e_t(M) }{dt}   &&  =  j_t(M-1 \to M)  - \epsilon 
\label{eshell1dib}
\end{eqnarray}

In terms of the velocities, Eq. \ref{eshell1di} corresponds for $1 \leq n \leq M-1$ to
\begin{eqnarray}
 \frac{d v_t(n)  }{dt}  && =Q_{\alpha,\beta} [v_t(n-1); v_t(n);v_t(n+1) ]  
\label{vshell1di}
\end{eqnarray}
while Eqs \ref{eshell1dib} yield the boundary dynamics at $n=0$ and $n=M$
\begin{eqnarray}
 \frac{d v_t(0) }{dt}   &&  =  \frac{ \epsilon }{v_t(0) } + Q_{\alpha,\beta} [0; v_t(0);v_t(1) ]  
\nonumber \\
 \frac{d v_t(M) }{dt}   &&  = Q_{\alpha,\beta} [v_t(M-1); v_t(M);0]    - \frac{ \epsilon }{v_t(M)  }
\label{vshell1dib}
\end{eqnarray}

A small perturbation $ u_t (n) $ around the time-independent solution of Eq. \ref{vn41}
\begin{eqnarray}
 v_t (n)  && =  v(n) + u_t (n)
\label{vper}
\end{eqnarray}
then follows the linearized Dynamics of Eq. \ref{vshell1di}
for $1 \leq n \leq M-1$
\begin{eqnarray}
  \frac{d  u_t (n) }{dt} && = G_{n,n-1} u_t(n-1) +   G_{n,n} u_t (n)  +  G_{n,n+1} u_t(n) 
\label{udyn}
\end{eqnarray}
with the boundary dynamics at $n=0$ and $n=M$ from Eqs \ref{eshell1dib} 
\begin{eqnarray}
 \frac{d u_t(0) }{dt}   &&  = \left[ - \frac{ \epsilon }{v^2(0) }  +  G_{0,0} \right] u_t (0)  +  G_{0,1} u_t(1) 
\nonumber \\
 \frac{d u_t(M) }{dt}   &&  =  G_{M,M-1} u_t(M-1) + \left[  G_{M,M}    + \frac{ \epsilon }{v^2(M)  } \right] u_t(M)
\label{ushell1dib}
\end{eqnarray}
where we have introduced the tridiagonal matrix elements
\begin{eqnarray}
G_{n,n-1} && \equiv  \frac{ \partial Q_{\alpha,\beta} [v(n-1); v(n);v(n+1) ]   }{ \partial v(n-1) } = \alpha  \frac{  2 v(n-1)   }{l_n}  + \beta \frac{v(n)  }{l_{n-1}} 
\nonumber \\
 G_{n,n} && \equiv \frac{ \partial Q_{\alpha,\beta} [v(n-1); v(n);v(n+1) ]   }{ \partial v(n) } =- \alpha  \frac{    v(n+1) }{l_{n+1}} +    \beta \frac{v(n-1)   }{l_{n-1}}  
\nonumber \\
G_{n,n+1} && \equiv \frac{ \partial Q_{\alpha,\beta} [v(n-1); v(n);v(n+1) ]   }{ \partial v(n+1) } = -   \alpha   \frac{  v(n)   }{l_{n+1}}  -  \beta \frac{2 v(n+1)  }{l_{n}}  
\label{matrixM}
\end{eqnarray}

One is interested into the $(M+1)$ eigenvalues $\lambda_m$ of this linearized system
in order to determine if they correspond to stable $Re(\lambda_m)<0$ or unstable $Re(\lambda_m)>0$ perturbations.
In particular, the sum of all eigenvalues is given by the sum of the diagonal elements
\begin{eqnarray}
 \sum_{m=0}^M \lambda_m 
&& = \sum_{n=0}^M G_{n,n} - \frac{ \epsilon }{v^2(0) }+ \frac{ \epsilon }{v^2(M)  } 
\nonumber \\
&&  = (\beta-\alpha ) \sum_{n=1}^{M-1}  \frac{v(n)}{l_n} + \left[ \beta \frac{v(0)  }{l_{0}}  - \frac{ \epsilon }{v^2(0) }\right]
+ \left[ - \alpha \frac{v(M)  }{l_{M}} + \frac{ \epsilon }{v^2(M)  } \right]
\label{trace}
\end{eqnarray}
The bulk contribution $1 \leq n \leq M-1$ corresponds as it should to the phase-space-volume expansion of Eq. \ref{divjphasespace}
evaluated at the steady-state $v(n)$ of Eq. \ref{vn41}.
The sum grows exponentially with the number $M$ of generations of the inertial range
\begin{eqnarray}
 (\beta-\alpha ) \sum_{n=1}^{M-1}  \frac{v(n)}{l_n} =  (\beta-\alpha )  \frac{v(0)  }{l_0 } \sum_{n=1}^{M-1} 2^{n (1-h) }
=  (\beta-\alpha )  \frac{v(0)  }{l_0 } \left(  \frac{ 2^{M (1-h)} - 2^{ (1-h)} } { 2^{ (1-h) }- 1 } \right)
\label{sum}
\end{eqnarray}
The constraint $\beta=\alpha$ of Eq. \ref{betachoice} coming from the phase-space-volume conservation in the inertial range
implies that the sum of the $(M+1)$ eigenvalues is only of order $O(1)$ from the boundary terms,
so one expects that there will be unstable perturbations associated to eigenvalues $\lambda_m$ with positive real parts $Re(\lambda_m)>0$.
On the contrary, all the cases $\beta \ne \alpha$ will produce an artificial dilatation or contraction of the phase-space-volume
around the steady state $v(n)$ and will shift the sum of the $(M+1)$ eigenvalues towards a positive or negative value growing exponentially
with the number of generations. 

Since the K41 steady-state of Eq. \ref{vn41} is dynamically unstable, 
the Desnyanski-Novikov shell model displays non-trivial dynamical properties
including chaoticity, intermittency, multifractality and soliton-like pulses as described in detail in Ref \cite{dombre}
(note that some translation is needed between the present choice of coefficients $(\alpha,\beta)$ in Eq. \ref{qab}
and the coefficients $(\alpha,\beta)$ defined in Eq 3 of Ref  \cite{dombre}).
However the Desnyanski-Novikov shell model cannot describe spatial heterogeneities of the Richardson cascade : 
in the next section, we thus describe how to include these effects via the scale-spatial-tree generalization.

%%%%%%%%%%%%%%%%%%%%%%%%%%%%%%%%%%%%%%%%%%%%%%%%%

\section{ Dynamics for scalar velocities $v_t(i_1,..,i_n)$ on the scale-spatial-tree  }

\label{sec_tree}

As recalled in the Introduction, the GOY shell model 
has been then extended to a scale-spatial tree structure \cite{benzi_1p1,benzi_inter,benzi_ultra},
in order to take into account the physical picture of the Richardson cascade.
In this section, our goal is to apply this idea to the Desnyanski-Novikov shell model recalled in the previous section.

\subsection{ Tree structure of scale-spatial cells $(i_1,..,i_n)$ }

\label{sec_cells}

Here we wish to give a real-space interpretation of the length scales $l_n=l_02^{-n}$ of Eq. \ref{ln}.
The generation $n=0$ corresponds to the biggest length-scale $l_0$ and to the total volume $l_0^D$ in dimension $D=3$.
The generation $n=1$ corresponds to the second biggest length-scale $l_1= \frac{l_0}{2} $ and to the volume $l_1^D= l_0^D 2^{-D} $,
so that there are 
\begin{eqnarray}
b = \frac{l_0^d}{l_1^d}=2^D
\label{bdef}
\end{eqnarray}
independent cells of the generation $n=1$ that will be labelled by the index $i_1=1,2,..,b$.
Similarly, each volume $i_1$ of generation $n=1$ contains $b$ independent volumes of the generation $n=2$ that will be labelled by the supplementary index $i_2=1,2,..,b$, and so on.
At generation $n$, there are thus
\begin{eqnarray}
{\cal N}_n = b^n
\label{nn}
\end{eqnarray}
independent cells labelled by the $n$ integers $(i_1,..,i_n)$ with $i_k=1,..,b$ for $k=1,..,n$,
that represents the whole genealogy of ancestors.
So the 1-dimensional scale space $n$ of the previous section has now been extended into a scale-spatial tree of branching $b=2^D$
with a clear interpretation in real-space.
Accordingly, the number of modes has changed from the logarithmic expression $N = \frac{\ln \left( \frac{l_0}{l_N}\right)}{\ln 2}$ 
of the 1-dimensional scale-lattice $n$
into the volumic expression
\begin{eqnarray}
{\cal N}^{tot}(0,N) = \sum_{n=0}^N {\cal N}_n = \frac{ b^{N+1} -1} { b-1} \simeq b^N = 2^{D N} = \left( \frac{l_0}{l_N}\right)^D
\label{nmodes}
\end{eqnarray}
for the present scale-spatial tree structure.

\subsection{ Observables associated to each scale-spatial cell $(i_1,..,i_n)$ }

In terms of the usual velocity field $\vec v_t (\vec r)$ of fluid mechanics, the total kinetic energy contained in the volume $l_0^D$ reads
\begin{eqnarray}
{\cal E}^{tot}_t  && = \int_{l_0^D} d^D \vec r  \frac{ [ \vec v_t (\vec r) ]^2 }{2}
\label{etotv}
\end{eqnarray}

Here we wish to decompose this total energy according to the scale-spatial tree structure of cells as
\begin{eqnarray}
{\cal E}^{tot}_t  && = {\cal E}_t(0)
+ \sum_{n=1}^{+\infty} \sum_{i_1=1}^{b} ... \sum_{i_n=1}^{b} {\cal E}_t (i_1,..,i_n)
\label{etot}
\end{eqnarray}
where the energy ${\cal E}_t (i_1,..,i_n)$ associated to the motion of velocity $v_t (i_1,..,i_n) $
in the cell $(i_1,..,i_n)$ of volume $l_n^D$
reads
\begin{eqnarray}
{\cal E}_t (i_1,..,i_n) = l_n^D \frac{v_t^2(i_1,..,i_n) }{2}
\label{evtree}
\end{eqnarray}
while the energy $ {\cal E}_t(0)$ associated to the integral scale is given similarly by
\begin{eqnarray}
{\cal E}_t (0) = l_0^D \frac{v_t^2(0) }{2}
\label{evtree0}
\end{eqnarray}
In terms of the velocities, the total energy of Eq. \ref{etot} thus reads
\begin{eqnarray}
{\cal E}^{tot}_t  && =  l_0^D \frac{v_t^2(0) }{2}
+ \sum_{n=1}^{+\infty} \sum_{i_1=1}^{b} ... \sum_{i_n=1}^{b}  l_n^D \frac{v_t^2(i_1,..,i_n) }{2}
\nonumber \\
&& = l_0^D \left[ \frac{v_t^2(0) }{2}
+ \sum_{n=1}^{+\infty}  \left(  b^{-n} \sum_{i_1=1}^{b} ... \sum_{i_n=1}^{b}   \frac{v_t^2(i_1,..,i_n) }{2} \right)  \right]
\label{etotvn}
\end{eqnarray}

\subsection{ Energy dynamics on the tree structure }

In the inertial range corresponding to generations $1 \leq n \ll N$,
where the energy is neither created nor destroyed, the Dynamics for the energy corresponds to
some discrete continuity equation.
In the present hierarchical tree structure, it is natural to assume that the cell $(i_1,..,i_n)$ is 
 able to exchange energy only with its ancestor $(i_1,..,i_{n-1})$ and with its $b$ children $(i_1,..,i_n,i_{n+1})$
with $i_{n+1}=1,2,..,b$. As a consequence, it will be convenient to denote $J_t(i_1,..,i_n)$
the energy-current received by the cell $(i_1,..,i_n)$ from its ancestor $(i_1,..,i_{n-1})$
in order to write the energy dynamics in the inertial range $1 \leq n \ll N$ as
\begin{eqnarray}
\frac{d  {\cal E}_t (i_1,..,i_n) }{dt} && = J_t (i_1,..i_n)  - \sum_{i_{n+1}=1}^b J_t (i_1,..,i_{n+1})
\label{energydyn}
\end{eqnarray}

For the generation $n=0$, there is no ancestor but the injected energy flow by the external forcing corresponds to
\begin{eqnarray}
J_t(0) \equiv \epsilon l_0^D
\label{j0injected}
\end{eqnarray}
so that the dynamical equation for ${\cal E}_t(0)$ reads
\begin{eqnarray}
\frac{d  {\cal E}_t(0) }{dt} && = \epsilon l_0^D - \sum_{i_1=1}^b J_t(i_1)
\label{energydyn0}
\end{eqnarray}

\subsection{ Energy current $J_t (i_1,..i_n) $   }

For the energy current received by the cell $(i_1,..,i_n)$ from its ancestor $(i_1,..,i_{n-1})$,
the form analogous to the 1-dimensional energy-current of Eq. \ref{jeshell1d} reads using Eq. \ref{evtree}
\begin{eqnarray}
J_t (i_1,..i_n ) &&  =  2 \alpha \frac{  {\cal E}_t(i_1,..,i_{n-1})}{b}  \frac{v_t(i_1,..,i_{n}) }{ l_n }  
+  2 \beta  {\cal E}_t(i_1,..,i_{n}) \frac{v_t(i_1,..,i_{n-1}) }{l_{n-1} }  
\nonumber \\
&& =  l_{n}^D  \left[ \alpha \frac{  v_t^2(i_1,..,i_{n-1})  v_t(i_1,..,i_n) }{l_n} 
 +   \beta  \frac{v_t(i_1,..,i_{n-1}) v_t^2(i_1,..,i_{n})  }{l_{n-1}}  \right]
\label{currentjadv}
\end{eqnarray}
so that the total energy current from the generation $(n-1)$ to the generation $n$ reads
\begin{eqnarray}
J_t ( n-1 \to n ) && \equiv \sum_{i_1=1}^b ...  \sum_{i_{n-1}=1}^b  \sum_{i_n=1}^b J_t (i_1,..i_n) 
\nonumber \\
&& = l_0^D    \frac{1}{b^n} \sum_{i_1=1}^b ...  \sum_{i_{n-1}=1}^b  \sum_{i_n=1}^b
 \left[   \alpha  \frac{  v_t^2(i_1,..,i_{n-1})  v_t(i_1,..,i_n) }{l_n} +\beta \frac{v_t(i_1,..,i_{n-1}) v_t^2(i_1,..,i_{n})  }{l_{n-1}}  \right]
\label{currentjadvtot}
\end{eqnarray}

\subsection{ Dynamical model for the velocities $v_t(i_1,..,i_n) $  }

In summary, putting everything together, 
the energy Dynamics of Eqs \ref{energydyn} \ref{energydyn0} 
involving the energy currents of Eq. \ref{currentjadv}
can be translated for the velocities of Eq. \ref{evtree}
into the following equations
\begin{eqnarray}
  \frac{d  v_t (i_1,..,i_n) }{dt} && = 
 \alpha  \frac{  v_t^2(i_1,..,i_{n-1})   }{l_n} 
 +  \beta \frac{v_t(i_1,..,i_{n-1}) v_t(i_1,..,i_{n})  }{l_{n-1}}  
\nonumber \\
&& 
- \frac{1}{b} \sum_{i_{n+1}=1}^b 
 \left[ \alpha  \frac{  v_t(i_1,..,i_{n})  v_t(i_1,..,i_{n+1}) }{l_{n+1}}  + \beta \frac{ v_t^2(i_1,..,i_{n+1})  }{l_{n}}  
  \right]
-  \frac{\nu D }{l_n^2}  v_t(i_1,..,i_n)
\label{vdyn}
\end{eqnarray}
with the boundary equation  for $n=0$ 
\begin{eqnarray}
\frac{d  v_t(0) }{dt} && = \frac{ \epsilon }{v_t(0) }
- \frac{1}{b}\sum_{i_{1}=1}^b 
\left[  \alpha  \frac{  v_t(0)  v_t(i_1) }{l_{1}} +   \beta  \frac{ v_t^2(i_1)  }{l_{0}}  \right]
-  \frac{\nu D }{l_0^2} v_t(0)
\label{vdyn0}
\end{eqnarray}

\subsection{ Link with the Desnyanski-Novikov 1-dimensional shell model }

If the initial velocities depend only on the generation index $n$
\begin{eqnarray}
v^{homo}_{t=0}(i_1,..,i_n) && =  v_{t=0}(n)
\label{homo0}
\end{eqnarray}
then this property is preserved by the dynamics at all times $t \geq 0$ 
\begin{eqnarray}
v^{homo}_{t}(i_1,..,i_n) && =  v_{t}(n)
\label{homo}
\end{eqnarray}
and the tree dynamical model of Eqs \ref{vdyn} then reduces to Desnyanski-Novikov 1-dimensional shell model 
recalled in section \ref{sec_DN}.

Note that the total energy of Eq. \ref{etotv} reduces to
\begin{eqnarray}
 {\cal E}^{tot}_t  && 
= l_0^D \sum_{n=0}^{+\infty}  \frac{v_t^2(n) }{2}  
\label{etothomo}
\end{eqnarray}
while the total energy current from the generation $(n-1)$ to the generation $n$ of Eq. \ref{currentjadvtot} becomes
\begin{eqnarray}
J_t ( n-1 \to n ) 
&& = l_0^D    
 \left[   \alpha  \frac{  v_t^2(n-1)  v_t(n) }{l_n}  +  \beta  \frac{v_t(n-1) v_t^2(n)  }{l_{n-1}}    \right]
\label{currentjadvtothomo}
\end{eqnarray}
i.e. one recovers the energy of Eq. \ref{e1d}
and the energy-current of Eq. \ref{jeshell1d}
up to the total volume factor $l_0^D$.

In conclusion, with respect to the Desnyanski-Novikov 1-dimensional shell model,
the tree dynamical model of Eqs \ref{vdyn} allows dynamical heterogeneities between
the tree branches, i.e. between the coherent structures of the same scale that are localized in different regions of the whole volume.

%%%%%%%%%%%%%%%%%%%%%%%%%%%%%%%%%%%%%%%%%%%%%%%%%%

\newpage

\section{ Dynamics for $3D$ velocity vectors $\vec v_t(n)$ on the 1-dimensional scale lattice $n$}

\label{sec_3D}

In this section, we wish to promote the scalar velocities $v_t(n)$ of the Desnyanski-Novikov 
described in section \ref{sec_DN} into $3D$ velocity vectors $\vec v_t(n)$.

\subsection{ Observables associated to the 3D-vector nature of the velocities $\vec v_t(n) $   }

\label{sec_vec}

It will be convenient to introduce the vector $\vec 1=(1,1,1)$ where the three components take the value unity.
At the length scale $l_n$, the gradient operator will be replaced by the vector
\begin{eqnarray}
(\vec \nabla)_n \to  \frac{ \vec 1}{l_n}
\label{grad}
\end{eqnarray} 

The scalar product with the velocity $\vec v_t(n) $ will produce the advection factor $a_t(n)$ at scale $l_n$
\begin{eqnarray}
(\vec v . \vec \nabla)_n \to  \frac{\vec v_t(n) . \vec 1}{l_n} = \sum_{\mu=x,y,z} \frac{v_t^{\mu}(n) }{l_n} \equiv a_t(n)
\label{atn}
\end{eqnarray} 

The vorticity $\omega = \vec \nabla \times \vec v$ will be defined at scale $l_n$ by
\begin{eqnarray}
(\vec \nabla \times \vec v)_n \to  \frac{ \vec 1 \times \vec v_t(n) }{l_n}  \equiv \vec \omega_t(n)
\label{omegatn}
\end{eqnarray} 
i.e. more explicitly for the 3D coordinates $\mu=x,y,z$
\begin{eqnarray}
 \omega_t^x(n) && \equiv \frac{ v_t^{z}(n) - v_t^{y}(n)}{l_n}
\nonumber \\
 \omega_t^y(n) && \equiv \frac{ v_t^{x}(n) - v_t^{z}(n)}{l_n}
\nonumber \\
 \omega_t^z(n) && \equiv \frac{ v_t^{y}(n) - v_t^{x}(n)}{l_n}
\label{omegaxyz}
\end{eqnarray}
So in this framework, the vorticity simply measures the differences between velocities components at scale $l_n$.
In particular, the sum of the three components of $\vec \omega_t(n)$ vanishes, so that there are only two independent components.
The velocity $\vec v_t(n)$ can be decomposed into the advection advection factor $a_t(n)$ of Eq \ref{atn}
 and the vorticity $\vec \omega_t(n)$ of Eq. \ref{omegatn} as
\begin{eqnarray}
\vec v(n) =\frac{ l_n }{3} \left[ a(n)  \vec 1 - \vec 1 \times \vec \omega(n) \right]
\label{vinversion}
\end{eqnarray} 
so that its square reads
\begin{eqnarray}
\vec v^2(n) = \frac{ l_n^2 }{3} \left[ a^2(n)  +\vec \omega^2(n) \right]
\label{v2inversion}
\end{eqnarray}

The angular momentum associated to the vorticity $\vec \omega_t(n) $ on the scale $l_n$ will be defined 
with a moment of inertia given by $l_n^2$ (discrete analog of the continuous relation \ref{omegar0})
\begin{eqnarray}
\vec L_t(n) \equiv l_n^2 \vec \omega_t(n)
\label{angularn}
\end{eqnarray} 

In this 3D version of the shell model, we will require that in the absence of external forcing and viscosity,
the conserved quantities (see section \ref{app_conserved} in Appendix \ref{app_navier}) are the total energy
\begin{eqnarray}
e_t^{tot} \equiv \sum_{n=0}^{+\infty} \frac{ \vec v_t^2(n)}{2}
\label{etotvec}
\end{eqnarray} 
and the total angular momentum
\begin{eqnarray}
\vec L_t^{tot} \equiv \sum_{n=0}^{+\infty} \vec L_t(n) =\sum_{n=0}^{+\infty} l_n^2 \vec \omega_t(n)
\label{angulartot}
\end{eqnarray} 
whose decomposition over the scales can be seen as an iterated version of Koenig Theorem for angular momentum.

\subsection{ Dynamics for the $3D$ velocity vectors $\vec v_t(n)$ }

We wish to generalize the scalar-velocity Dynamics of Eq. \ref{vdyn} into an equation for the 3D velocity vector $\vec v_t(n)$
\begin{eqnarray}
 \frac{d \vec v_t(n)  }{dt}  && = \vec Q(\vec v_t(n-1) ,\vec v_t(n),\vec v_t(n+1)    )  -  \frac{\nu D }{l_n^2} \vec v_t(n)
\label{vdynvec}
\end{eqnarray}
where the boundary equation for $n=0$ includes the external forcing $\vec f [ \vec v_t(0)]  $ 
\begin{eqnarray}
 \frac{d \vec v_t(0)  }{dt}  && = \vec Q(0 ,\vec v_t(0),\vec v_t(1)    ) + \vec f [ \vec v_t(0)] -  \frac{\nu D }{l_0^2} \vec v_t(0)
\label{vdynvec0}
\end{eqnarray}

The nonlinear quadratic term $\vec Q $ describing the conservative part of the dynamics
should conserve the total energy of Eq. \ref{etot}, the total angular momentum of Eq. \ref{angulartot},
and the phase-space-volume.
The simplest choice 
that mimics the scalar-velocity Dynamics of Eq. \ref{vdyn} for $\alpha=\beta=1$ 
and the 3D Navier Equation of Eq. \ref{navierlamb}
 reads
\begin{eqnarray}
\vec Q(\vec v_t(n-1) ,\vec v_t(n),\vec v_t(n+1)    )
&& =   \left[    a_t(n-1)  -   a_t(n+1) \right]  \vec v_t(n)
 + \frac{ \vec 1}{l_n}
\left[  \vec v^2_t(n-1)  - \vec v^2_t(n+1) 
\right]
\nonumber \\
&& + \left[ 4 \vec \omega_t(n-1) -  \frac{1}{4}  \vec \omega_t(n+1)\right] \times \vec v_t(n) 
\label{vecqnlamb}
\end{eqnarray}
The first term containing the advection factors introduced in Eq. \ref{atn} reads more explicitly in terms of the velocity components
\begin{eqnarray}
  \left[    a_t(n-1)  -   a_t(n+1) \right]  \vec v_t(n) =   \left[    \sum_{\mu=x,y,z} \frac{v_t^{\mu}(n-1) }{l_{n-1}}   -  \sum_{\mu=x,y,z} \frac{v_t^{\mu}(n+1) }{l_{n+1}}  \right]  \vec v_t(n) 
\label{qadvection}
\end{eqnarray}
and generalizes the first term of Eq. \ref{qaa}.
The second term in Eq. \ref{vecqnlamb} involves the total kinetic energy of the neighboring levels and is also a direct generalization
of the second term of Eq. \ref{qaa}.
Finally the third term of Eq. \ref{vecqnlamb} is the complete novelty with respect to the scalar case of Eq. \ref{qaa}
 and mimics the Lamb vector of the Navier-Stokes Eq \ref{navierlamb}.
Using the vorticity introduced in Eq. \ref{omegatn}, one obtains the identities
\begin{eqnarray}
  \vec \omega_t(n-1) \times \vec v_t(n) 
&& =\frac{ 1 }{2}  \left[ \vec v_t(n-1)  a_t(n) -  \frac{  \vec 1  }{ l_{n}} ( \vec v_t(n-1). \vec v_t(n) ) \right]
\nonumber \\
  \vec \omega_t(n+1) \times \vec v_t(n) 
&& =2 \left[  \vec v_t(n+1)   a_t(n) -   \frac{  \vec 1  }{l_{n}} ( \vec v_t(n+1). \vec v_t(n) ) \right]
\label{omrotvpq}
\end{eqnarray} 

Putting everything together, the dynamical equation for the component $v^{\mu}_t(n)$ reads explicitly in terms of the other velocity components alone
\begin{eqnarray}
 && Q^{\mu}(\vec v_t(n-1) ,\vec v_t(n),\vec v_t(n+1)    )
 =   \left[   \sum_{\mu'} \frac{v_t^{\mu'}(n-1) }{l_{n-1}}   -  \sum_{\mu'} \frac{v_t^{\mu'}(n+1) }{l_{n+1}} \right]  v_t^{\mu}(n)
  \nonumber \\ &&
 + \frac{  1}{l_n}  \left(   \sum_{\mu'} [ v^{\mu'}_t(n-1)]^2  - \sum_{\mu'} [ v^{\mu'}_t(n+1)]^2   \right)
\nonumber \\
&& +  \left[ 2  v_t^{\mu}(n-1) -  \frac{1}{2}   v^{\mu}_t(n+1)     \right]  \sum_{\mu'} \frac{v_t^{\mu'}(n) }{l_n}
+ \frac{   1  }{ l_{n}} \sum_{\mu'} \left[ - 2  v^{\mu'}_t(n-1)  + \frac{1}{2}   v^{\mu'}_t(n+1)   \right] v^{\mu'}_t(n) ) 
\label{vecqnlambmu}
\end{eqnarray}

 Note that the local phase-space dilatation for the component $v^{\mu}_t(n) $ reduces to the advection factors
\begin{eqnarray}
&& \frac{\partial  Q^{\mu}_n(\vec v_t(n-1) ,\vec v_t(n),\vec v_t(n+1)    ) }{ \partial v_t^{\mu}(n) }
 =    \sum_{\mu'} \frac{v_t^{\mu'}(n-1) }{l_{n-1}}   -  \sum_{\mu'} \frac{v_t^{\mu'}(n+1) }{l_{n+1}} 
= a_t(n-1)- a_t(n+1)
\label{volmu}
\end{eqnarray}
so that the total divergence of the phase-space-current vanishes in the bulk concerning the inertial range as in Eq. \ref{divjphasespace}
\begin{eqnarray}
 \sum_{n}   \sum_{\mu} \frac{\partial  Q^{\mu}_n(\vec v_t(n-1) ,\vec v_t(n),\vec v_t(n+1)    ) }{ \partial v_t^{\mu}(n) } =0
\label{divjphasespacevec}
\end{eqnarray}

\subsection{ Dynamics for the kinetic energy}

The dynamical equation of the kinetic energy
\begin{eqnarray}
e_t(n) \equiv \frac{ \vec v_t^2(n)}{2}
\label{en}
\end{eqnarray} 
is obtained from the scalar product of Eq. \ref{vdynvec} with the velocity $\vec v_t(n)$,
so that the contributions of the Lamb vector in Eq. \ref{vecqnlamb}
 vanishes and one obtains

\begin{eqnarray}
 \frac{d e_t(n) }{dt} =  \vec v_t(n) .\frac{d \vec v_t(n) }{dt} 
= j_t(n-1 \to n)   - j_t(n \to n+1)  -  \frac{\nu D }{l_n^2} \vec v^2_{n}(t)
\label{energydynvec}
\end{eqnarray}
in terms of the energy current 
 from generation $(n-1)$ to generation $n$
\begin{eqnarray}
j_t(n-1 \to n)  \equiv   \vec v_t^2(n-1) a_t(n)  +  \vec  v^2_t(n) a_t(n-1) 
= \vec v_t^2(n-1)  \left( \sum_{\mu} \frac{v_t^{\mu}(n) }{l_n}  \right) +  \vec  v^2_t(n) \left( \sum_{\mu} \frac{v_t^{\mu}(n-1) }{l_{n-1}} \right)
\label{jevec}
\end{eqnarray}
that directly generalizes the scalar expression of Eq. \ref{jeshell1d}, with kinetic energies and advection factors of the two shells.

\subsection{ Dynamics for the vorticity  }

The dynamical equation for the vorticity introduced in Eq. \ref{omegaxyz} is obtained from Eq. \ref{vdynvec} 
and Eq. \ref{vecqnlambmu}
\begin{eqnarray}
 \frac{d \vec \omega_t(n)  }{dt}  && = 
 \left[    a_t(n-1)  -   a_t(n+1) \right]  \vec \omega_t(n) 
+ \left[ 4 \vec \omega_t (n-1)  - \frac{1}{4}  \vec \omega_t(n+1) \right] a_t(n) 
 -  \frac{\nu D }{l_n^2} \vec \omega_t(n)
\label{omegadyn}
\end{eqnarray}
As explained in sections \ref{sec_omega} \ref{sec_L} of the Appendix,
the interpretation of the Dynamics for the vorticity in shell models
is clearer via the Dynamics of the eddy angular momentum of Eq. \ref{angularn}.

\subsection{ Dynamics for the angular momentum  }

The dynamical equation for the angular momentum introduced in Eq. \ref{angularn} is obtained from Eq. \ref{omegadyn} 
using $l_{n-1}= 2 l_n$ 
\begin{eqnarray}
 \frac{d \vec L_t(n)  }{dt} && = l_n^2  \frac{d \vec \omega_t(n)  }{dt}  
 =\vec K_t(n-1 \to n)   - \vec K_t(n \to n+1)  
 - \frac{\nu D }{l_n^2} \vec L_t(n)
\label{angulardyn}
\end{eqnarray}
in terms of the angular momentum current 
 from generation $(n-1)$ to generation $n$
\begin{eqnarray}
\vec K_t(n-1 \to n)  \equiv   \vec L_t (n-1) a_t(n)  +   \vec L_t(n) a_t(n-1) 
=   \vec L_t (n-1)   \left( \sum_{\mu} \frac{v_t^{\mu}(n) }{l_n}  \right) +  \vec L_t(n) \left( \sum_{\mu} \frac{v_t^{\mu}(n-1) }{l_{n-1}} \right)
\label{angulark}
\end{eqnarray}
whose structure in terms of the advection factors is completely similar to the energy current of Eq. \ref{jeshell1d}.

The corresponding cascade of angular momentum from large scales to small scales that is coupled with the energy cascade 
will produce new properties with respect to the scalar model that contains only the energy cascade.

\subsection{ Link with the Desnyanski-Novikov scalar shell model }

If the initial vector velocities are of the form
\begin{eqnarray}
\vec v_{t=0}(n) =  v_{t=0}(n)  \frac{ \vec 1 }{D} 
\end{eqnarray}
then this property is preserved by the dynamics in the absence of external forcing
\begin{eqnarray}
\vec v_t(n) =  v_t(n)  \frac{ \vec 1 }{D} 
\label{vecscal}
\end{eqnarray}
Since the components $\mu=x,y,z$ coincide, the vorticities identically vanish
\begin{eqnarray}
 \vec \omega_t(n) =0
\label{omegava}
\end{eqnarray}
while the advection factors of Eq. \ref{atn} and the kinetic energies simplify into
\begin{eqnarray}
 a_t(n) && = \sum_{\mu} \frac{v_t^{\mu}(n) }{l_n} =  \frac{v_t(n) }{l_n}
\nonumber \\
\vec v_t^2(n) && =  v_t^2(n)  \frac{  1 }{D} 
\label{atns}
\end{eqnarray} 
so that the Dynamics of Eq. \ref{vecqnlamb} for the velocity vector of Eq. \ref{vecscal}
reduces to Eq. \ref{qaa} for the scalar velocity $v_t(n)$.

In conclusion, with respect to the Desnyanski-Novikov 1-dimensional shell model,
the vectorial dynamical model of Eqs \ref{vdynvec} allows to allow heterogeneities between
the three components of the velocities and to introduce vorticity effects in the dynamics.
In the next section, we describe how to add spatial heterogeneities via the scale-spatial-tree structure.

\section{ Dynamics for $3D$ velocity vectors $\vec v_t(i_1,..,i_n)$ on the scale-spatial-tree }

\label{sec_3Dtree}

In this section, we put together the two generalizations introduced in the two previous sections
in order to obtain the dynamical model for $3D$ velocity vectors $\vec v_t(i_1,..,i_n)$ on the scale-spatial-tree.

\subsection{ Observables associated to each scale-spatial cell $(i_1,..,i_n)$ }

It is straightforward to generalize the vectorial observables introduced in section \ref{sec_vec}
to the tree structure described in \ref{sec_cells}.
The advection factor associated to the 3D vector velocity $\vec v_t(i_1,..,i_n)$ is
\begin{eqnarray}
 a_t(i_1,..,i_n) \equiv \frac{\vec v_t(i_1,..,i_n) . \vec 1}{l_n} = \sum_{\mu=x,y,z} \frac{v_t^{\mu}(i_1,..,i_n) }{l_n} 
\label{atntree}
\end{eqnarray} 
while the vorticity is defined by
\begin{eqnarray}
\vec \omega_t(i_1,..,i_n) \equiv \frac{ \vec 1 \times \vec v_t(i_1,..,i_n) }{l_n}  
\label{omegatntree}
\end{eqnarray} 
i.e. more explicitly for the 3D coordinates $\mu=x,y,z$
\begin{eqnarray}
 \omega_t^x(i_1,..,i_n) && \equiv \frac{ v_t^{z}(i_1,..,i_n) - v_t^{y}(i_1,..,i_n)}{l_n}
\nonumber \\
 \omega_t^y(i_1,..,i_n) && \equiv \frac{ v_t^{x}(i_1,..,i_n) - v_t^{z}(i_1,..,i_n)}{l_n}
\nonumber \\
 \omega_t^z(i_1,..,i_n) && \equiv \frac{ v_t^{y}(i_1,..,i_n) - v_t^{x}(i_1,..,i_n)}{l_n}
\label{omegaxyztree}
\end{eqnarray}

\subsection{ Dynamics for the velocities }

The dynamical equations for the velocities that generalize both Eqs \ref{vdyn}
and \ref{vecqnlamb}
read
\begin{eqnarray}
  \frac{d \vec v_t (i_1,..,i_n) }{dt} && =   \left[    a_t(i_1,..,i_{n-1})  -    \frac{1}{b} \sum_{i_{n+1}=1}^b  a_t(i_1,..,i_{n+1}) \right]  \vec v_t (i_1,..,i_n)
 \nonumber \\
&&+ 
\left[  \vec v^2_t(i_1,..,i_{n-1})  -   \frac{1}{b} \sum_{i_{n+1}=1}^b \vec v^2_t(i_1,..,i_{n+1})   \right] \frac{ \vec 1}{l_n}
\nonumber \\
&& + \left[ 4 \vec \omega_t(i_1,..,i_{n-1}) -  \frac{1}{4 b } \sum_{i_{n+1}=1}^b \vec \omega_t(i_1,..,i_{n+1})\right] \times \vec v_t(i_1,..,i_n) 
 -  \frac{\nu D }{l_n^2} \vec v_t(i_1,..,i_n)
\label{vvectree}
\end{eqnarray}
with the boundary dynamics at $n=0$ that includes some external forcing $ \vec f [ \vec v_t (0) ] $
\begin{eqnarray}
  \frac{d \vec v_t (0) }{dt} && =  -   \left[        \frac{1}{b} \sum_{i_{1}=1}^b  a_t(i_1) \right]  \vec v_t (0)
 -  \left[      \frac{1}{b} \sum_{i_{1}=1}^b \vec v^2_t(i_1)   \right] \frac{ \vec 1}{l_0}
 -  \left[    \frac{1}{4 b } \sum_{i_{1}=1}^b \vec \omega_t(i_1)\right] \times \vec v_t(0) 
 -  \frac{\nu D }{l_0^2} \vec v_t(0) + \vec f [ \vec v_t (0) ]
\label{vvectree0}
\end{eqnarray}

\subsection{ Dynamics for the energies }

The scalar product of Eq. \ref{vvectree} with $v_t (i_1,..,i_n) $ yields the dynamics for the kinetic energy $ \frac{ \vec v^2_t (i_1,..,i_n)}{2} $ 
\begin{eqnarray}
  \frac{d  }{dt} \left( \frac{ \vec v^2_t (i_1,..,i_n)}{2} \right)&& = 
  \left[    a_t(i_1,..,i_{n-1})  -    \frac{1}{b} \sum_{i_{n+1}=1}^b  a_t(i_1,..,i_{n+1}) \right]  \vec v^2_t (i_1,..,i_n)
 \nonumber \\
&&+ 
\left[  \vec v^2_t(i_1,..,i_{n-1})  -   \frac{1}{b} \sum_{i_{n+1}=1}^b \vec v^2_t(i_1,..,i_{n+1})   \right] a_t(i_1,..,i_{n}) 
 -  \frac{\nu D }{l_n^2} \vec v^2_t(i_1,..,i_n)
\label{evectree}
\end{eqnarray}
with the boundary dynamics at $n=0$ that includes the power $\vec v_t (0) .  \vec f [ \vec v_t (0) ]  $  injected by the external forcing 
\begin{eqnarray}
 \frac{d  }{dt} \left( \frac{ \vec v^2_t (0)}{2} \right)&& =   
-   \left[        \frac{1}{b} \sum_{i_{1}=1}^b  a_t(i_1) \right]  \vec v^2_t (0)
 -  \left[      \frac{1}{b} \sum_{i_{1}=1}^b \vec v^2_t(i_1)   \right] a_t(0)
 -  \frac{\nu D }{l_0^2} \vec v^2_t(0) + \vec v_t (0) .  \vec f [ \vec v_t (0) ] 
\label{evectree0}
\end{eqnarray}

For the volumic energies generalizing Eq. \ref{evtree}
\begin{eqnarray}
{\cal E}_t (i_1,..,i_n) = l_n^D \frac{ \vec v_t^2(i_1,..,i_n) }{2}
\label{evtreevec}
\end{eqnarray}
the dynamical equation
\begin{eqnarray}
  \frac{d {\cal E}_t (i_1,..,i_n)  }{dt} && = J_t (i_1,..i_n ) - \sum_{i_{n+1}=1}^b  J_t (i_1,..i_{n+1} ) 
 -  \frac{\nu D }{l_n^2} 2 {\cal E}_t(i_1,..,i_n)
\label{volevectree}
\end{eqnarray}
involves the energy current that generalizes Eq. \ref{currentjadv}
\begin{eqnarray}
J_t (i_1,..i_n ) &&  =  2  \frac{  {\cal E}_t(i_1,..,i_{n-1})}{b}  a_t(i_1,..,i_{n})
+  2   {\cal E}_t(i_1,..,i_{n}) a_t(i_1,..,i_{n-1})
\nonumber \\
&& =  l_{n}^D \left[ \vec v_t^2(i_1,..,i_{n-1}) a_t(i_1,..,i_n) 
 +  \vec  v_t^2(i_1,..,i_{n}) a_t(i_1,..,i_{n-1}) \right]
\label{currentjadvv}
\end{eqnarray}
while the dynamics at generation $n=0$  
\begin{eqnarray}
  \frac{d {\cal E}_t (0)  }{dt} && =  - \sum_{i_{1}=1}^b  J_t (i_1 ) 
 -  \frac{\nu D }{l_0^2} 2 {\cal E}_t(0) + l_0^D  \vec v_t (0) .  \vec f [ \vec v_t (0) ] 
\label{volevectree0}
\end{eqnarray}
includes the total power $ l_0^D \vec v_t (0) .  \vec f [ \vec v_t (0) ]  $  injected by the external forcing
that is usually taken to be $l_0^D \epsilon$.

\subsection{ Dynamics for the vorticities }

The dynamics for the vorticity defined in Eq. \ref{omegatntree} is obtained by applying $\left( \frac{ \vec 1  }{l_n} \times  \right)$ to Eq. \ref{vvectree}
\begin{eqnarray}
  \frac{d \vec \omega_t (i_1,..,i_n) }{dt} && =  
 \left[    a_t(i_1,..,i_{n-1})  -    \frac{1}{b} \sum_{i_{n+1}=1}^b  a_t(i_1,..,i_{n+1}) \right]  \vec \omega_t (i_1,..,i_n)
\nonumber \\
&& + \left[ 4 \vec \omega_t(i_1,..,i_{n-1}) -  \frac{1}{4 b } \sum_{i_{n+1}=1}^b \vec \omega_t(i_1,..,i_{n+1})\right] a_t(n)
 -  \frac{\nu D }{l_n^2} \vec \omega_t(i_1,..,i_n)
\label{omegavectree}
\end{eqnarray}
with the boundary dynamics at $n=0$ 
\begin{eqnarray}
  \frac{d \vec \omega_t (0) }{dt} && =
-   \left[        \frac{1}{b} \sum_{i_{1}=1}^b  a_t(i_1) \right]  \vec \omega_t (0)
 -  \left[    \frac{1}{4 b } \sum_{i_{1}=1}^b \vec \omega_t(i_1)\right] a_t(0) 
 -  \frac{\nu D }{l_0^2} \vec \omega_t(0) + \frac{ \vec 1}{l_0} \times \vec f [ \vec v_t (0) ] 
\label{omegavectree0}
\end{eqnarray}

\subsection{ Dynamics for the angular momentum }

As for the total energy in Eq. \ref{etotv}, the total angular momentum in the volume $l_0^D$ reads
in terms of the velocity field $\vec v_t (\vec r)$ of fluid mechanics
\begin{eqnarray}
\vec{\cal L}^{tot}_t  && = \int_{l_0^D} d^D \vec r  \ [ \vec r \times  \vec v_t (\vec r)  ]
\label{ltotv}
\end{eqnarray}
The decomposition into the scale-spatial tree structure reads as in Eq. \ref{etot}
\begin{eqnarray}
\vec {\cal L}^{tot}_t  && = {\cal L}_t(0)
+ \sum_{n=1}^{+\infty} \sum_{i_1=1}^{b} ... \sum_{i_n=1}^{b} \vec {\cal L}_t (i_1,..,i_n)
\label{ltot}
\end{eqnarray}
where the contribution $\vec {\cal L}_t (i_1,..,i_n) $ of the cell $(i_1,..,i_n)$
involves the angular momentum $\vec L_t (i_1,..,i_n) = l_n^2 \vec\omega_t (i_1,..,i_n) $ and the volume $l_n^D$ 
\begin{eqnarray}
\vec{\cal L}_t (i_1,..,i_n) = l_n^D \vec L_t(i_1,..,i_n) 
 =  l_n^{D} \left[   l_n^2  \vec\omega_t (i_1,..,i_n) \right]
\label{lvtree}
\end{eqnarray}
Eq. \ref{ltot} can be thus rewritten as
\begin{eqnarray}
\vec {\cal L}^{tot}_t  && =  l_0^D \vec L_t(0) 
+ \sum_{n=1}^{+\infty} \sum_{i_1=1}^{b} ... \sum_{i_n=1}^{b}  l_n^D \vec L_t(i_1,..,i_n) 
\nonumber \\
&& = l_0^D \left[ \vec L_t(0) 
+ \sum_{n=1}^{+\infty}  \left(  b^{-n} \sum_{i_1=1}^{b} ... \sum_{i_n=1}^{b} \vec L_t(i_1,..,i_n)    \right)  \right]
\nonumber \\
&& = l_0^D \left[ l_0^2 \vec\omega_t (0)
+ \sum_{n=1}^{+\infty}  \left(  b^{-n} \sum_{i_1=1}^{b} ... \sum_{i_n=1}^{b} l_n^2 \vec\omega_t (i_1,..,i_n)   \right)  \right]
\label{ltotvn}
\end{eqnarray}

From the Dynamics of the vorticity in Eq. \ref{omegavectree}, one obtains the Dynamics for $\vec L_t (i_1,..,i_n) = l_n^2 \vec\omega_t (i_1,..,i_n) $ 
\begin{eqnarray}
  \frac{d \vec L_t (i_1,..,i_n) }{dt} && =  
 \left[    a_t(i_1,..,i_{n-1})  -    \frac{1}{b} \sum_{i_{n+1}=1}^b  a_t(i_1,..,i_{n+1}) \right]  \vec L_t (i_1,..,i_n)
\nonumber \\
&& + \left[  \vec L_t(i_1,..,i_{n-1}) -  \frac{1}{ b } \sum_{i_{n+1}=1}^b \vec L_t(i_1,..,i_{n+1})\right] a_t(n)
 -  \frac{\nu D }{l_n^2} \vec L_t(i_1,..,i_n)
\label{dynlvectree}
\end{eqnarray}
The dynamics for the volumic contribution $ \vec{\cal L}_t (i_1,..,i_n)= l_n^D \vec L_t (i_1,..,i_n) $  
\begin{eqnarray}
  \frac{d \vec {\cal L}_t (i_1,..,i_n) }{dt} && =  \vec {\cal K}_t (i_1,..i_n)  - \sum_{i_{n+1}=1}^b \vec{ \cal K}_t (i_1,..,i_{n+1})
 -  \frac{\nu D }{l_n^2} \vec {\cal L}_t(i_1,..,i_n)
\label{dynvollvectree}
\end{eqnarray}
involves the angular-momentum-current received by the cell $(i_1,..,i_n)$ from its ancestor $(i_1,..,i_{n-1})$
\begin{eqnarray}
\vec {\cal K}_t (i_1,..,i_n)  && =  \frac{ \vec {\cal L}_t(i_1,..,i_{n-1}) }{b}  a_t(i_1,..,i_n) +    a_t(i_1,..,i_{n-1})    \vec {\cal L}_t (i_1,..,i_n)  
\nonumber \\
&& =  l_{n}^D \left[  \vec L_t(i_1,..,i_{n-1}) a_t(i_1,..,i_n) 
 +    \vec {\cal L}_t(i_1,..,i_{n}) a_t(i_1,..,i_{n-1}) \right]
\label{klcurrent}
\end{eqnarray}
of the same form as the energy-current of Eq. \ref{currentjadvv}.
The dynamics at generation $n=0$  
\begin{eqnarray}
  \frac{d \vec {\cal L}_t (0) }{dt} && =   - \sum_{i_{1}=1}^b \vec{ \cal K}_t (i_1)
 -  \frac{\nu D }{l_0^2} \vec {\cal L}_t(0) + l_0^D \left( l_0 \vec 1 \times \vec f [ \vec v_t (0) ] \right) 
\label{dynvollvectree0}
\end{eqnarray}
includes the total effective torque $  l_0^D \left( l_0 \vec 1 \times \vec f [ \vec v_t (0) ] \right)$  exerted by the external forcing.

%%%%%%%%%%%%%%%%%%%%%%%%%%%%%%%%%%%%%%%%%%%%%%%%%%%%

\section{ Conclusions }

\label{sec_conclusion}

In summary, we have first revisited the Desnyanski-Novikov shell model for scalar velocities $v_t(n)$
on the one-dimensional-lattice $n=0,1,2,..$ labelling the length-scales $l_n=l_0 2^{-n}$,
in order to stress the importance of the Liouville phase-space-volume conservation that fixes the relation $\alpha=\beta$
between the two parameters that are usually considered as independent.
We have then described in detail how this model can be generalized in two directions :

  (i) the one-dimensional-lattice $n=0,1,2,..$ labelling the length-scales $l_n=l_0 2^{-n}$
can be replaced by a scale-spatial tree structure of nested cells in order to allow spatial heterogeneities between different coherent structures
that are localized in different regions of the whole volume 

 (ii) the scalar velocities $v_t(n)$ can be replaced by 3D-vector velocities $\vec v_t(n)$ 
in order to take into account the vorticity $\vec \omega_t(n)$ in the dynamical equations and to include vortex-stretching effects
via the cascade of angular momentum in addition to the cascade of energy.

Along the paper, we have stressed the similarities and differences with the Navier-Stokes dynamics
for the usual fields of fluid mechanics recalled in Appendix \ref{app_navier}.

Further work is needed to characterize the dynamical properties of the final dynamical model
concerning 3D velocities on the scale-spatial tree structure, in order to compare
with the chaoticity, intermittency, multifractality and soliton-like pulses of the
 Desnyanski-Novikov shell model \cite{dombre}.

\appendix

\section{ Reminder on the Navier-Stokes dynamics  }

\label{app_navier}

In this Appendix, we recall the properties of the Navier-Stokes dynamics for various observables
in order to stress the similarities and differences with the shell models discussed in the text.

\subsection{ Navier-Stokes dynamics for the velocity  $\vec v_t(\vec r)$ }

For an incompressible fluid, the velocity field $v_t(\vec r)$ is divergence-free
\begin{eqnarray}
\vec \nabla . \vec v   =0 
\label{divzero}
\end{eqnarray} 
and evolves according to the Navier-Stokes equation
\begin{eqnarray}
\partial _t \vec v = - \left( \vec v . \vec \nabla \right) \vec v - \vec \nabla P  + \nu \Delta \vec v + \vec f
\label{navier}
\end{eqnarray} 
that involves the advection term, the gradient of the pressure $P$, the dissipation induced by the viscosity $\nu$ and the external force $\vec f$.

The vorticity field
\begin{eqnarray}
\vec \omega \equiv \vec \nabla \times \vec v
\label{vorticity}
\end{eqnarray} 
allows to rewrite the advection term via the identity
\begin{eqnarray}
\left( \vec v . \vec \nabla \right) \vec v  =\vec \nabla \left( \frac{ \vec v^2}{2} \right) + \vec \omega \times \vec v
\label{advectionidentity}
\end{eqnarray} 
and to obtain the following alternative form of the Navier-Stokes Equation \ref{navier}
\begin{eqnarray}
\partial _t \vec v = - \vec \nabla \left( \frac{ \vec v^2}{2} + P  \right)   - \vec \omega \times \vec v  + \nu \Delta \vec v + \vec f
\label{navierlamb}
\end{eqnarray} 
The first term corresponds to the gradient of the Bernoulli function $\left(\frac{ \vec v^2}{2} +P \right)$ containing the kinetic energy $ \frac{ \vec v^2}{2}  $
and the pressure $P$. The second term involving the Lamb vector $( \vec \omega \times \vec v )$ is analogous to some Coriolis force for the velocity field under the effect of its own rotation.

\subsection{ Dynamics for the kinetic energy  $\left( \frac{ \vec v_t^2(\vec r)}{2} \right) $ }

The dynamical equation for the kinetic energy $\left( \frac{ \vec v_t^2(\vec r)}{2} \right) $  is obtained from the scalar product of Eq. \ref{navierlamb} with the velocity $\vec v$,
so that the contribution of the Lamb vector vanishes $(\vec \omega \times \vec v). \vec v =0$ and one obtains
\begin{eqnarray}
\partial _t\left( \frac{ \vec v^2}{2} \right) = \vec v . \partial _t \vec v = 
- ( \vec v .\vec \nabla ) \left(  \frac{ \vec v^2}{2}   \right)  
-  \vec v .\vec \nabla P
+\nu \vec v .  \Delta \vec v  + \vec v . \vec f
\label{naviere}
\end{eqnarray} 
The first term corresponds to the advection of the energy, the third term corresponds to the dissipation by the viscosity,
while the second and fourth terms correspond to the power of the pressure gradient force and of the external force respectively.

\subsection{ Dynamics for the vorticity $\vec \omega_t (\vec r)$ }

\label{sec_omega}

The Dynamics for the vorticity $\vec \omega_t(\vec r) $ is obtained by applying the curl operator $(\vec \nabla \times )$ to Eq.  \ref{navierlamb}
so that the contribution of the gradient term of the Bernoulli function vanishes and one obtains
\begin{eqnarray}
\partial _t \vec \omega && = \vec  \nabla \times \left( \vec v \times \vec \omega \right) 
 + \nu \Delta \vec \omega + \vec  \nabla \times \vec f 
 \nonumber \\  &&
  = - (\vec v . \vec \nabla ) \vec \omega + (\vec \omega . \vec \nabla) \vec v
+ \nu \Delta \vec \omega+ \vec  \nabla \times \vec f 
\label{navierrot}
\end{eqnarray} 
 The first term represents the advection of the vorticity, the third term describes the dissipation by the viscosity,
while the fourth term represents the generation of vorticity by the external force.
The second term $ (\vec \omega . \vec \nabla) \vec v$ for the Dynamics of the component $\omega_x$ of the vorticity
can be rewritten as
\begin{eqnarray}
 (\vec \omega . \vec \nabla) v_x =  \omega_x  \partial_x  v_x +  (\omega_y  \partial_y + \omega_z  \partial_z ) v_x 
\label{stretching}
\end{eqnarray} 
where the first term describes the vortex stretching effect when $\omega_x  \partial_x  v_x >0 $, while the other terms correspond to vortex turning.
In turbulence, the vortex stretching is considered as one of the essential mechanism for the cascade towards smaller scales.

\subsection{ Dynamics for the angular momentum }

\label{sec_L}

Physically, it is clear that the vortex stretching effect mentioned above is related to the local conservation of angular momentum.
To have a more direct correspondence with the shell model described in section \ref{sec_3D},
it is actually useful to consider now the Dynamics of the angular momentum,
even if it is rather unusual in fluid mechanics.

The Dynamics of the angular momentum with respect to some point $\vec r_0 $
\begin{eqnarray}
\vec L_t(\vec r \vert \vec r_0) \equiv ( \vec r -  \vec r_0 ) \times \vec v_t (\vec r)
\label{angular}
\end{eqnarray} 
is obtained by applying the operator $( \vec r -  \vec r_0 )  \times $ to Eq. \ref{navier} and by rearranging the terms into
\begin{eqnarray}
\partial _t \vec L = - \left( \vec v . \vec \nabla \right) \vec L
 - ( \vec r -  \vec r_0 ) \times (\vec \nabla P ) + \nu ( \vec r -  \vec r_0 ) \times (\Delta \vec v ) +( \vec r -  \vec r_0 ) \times \vec f
\label{navierl}
\end{eqnarray} 
The first term corresponds to the advection, the third term corresponds to the dissipation by the viscosity,
while the second and fourth terms correspond to the torque of the pressure gradient force and of the external force respectively.

The link between the angular momentum of Eq. \ref{angular} and the vorticity $\vec \omega (\vec r)$ is given by the identity
concerning the Laplacian of the angular momentum
\begin{eqnarray}
\Delta \vec L_t(\vec r \vert \vec r_0) = ( \vec r -  \vec r_0 ) \times \Delta\vec v_t (\vec r) + 2 \vec \omega (\vec r)
\label{angularomega}
\end{eqnarray} 
i.e. the vorticity at $ \vec r_0$ can be interpreted as the limit of the Laplacian of the angular momentum $\vec L_t(\vec r \vert \vec r_0) $ as $\vec r \to \vec r_0$
\begin{eqnarray}
\vec \omega (\vec r_0)   =\frac{1}{2}  \lim_{\vec r \to \vec r_0} \left[  \Delta \vec L_t(\vec r \vert \vec r_0) \right]
\label{omegar0}
\end{eqnarray}

\subsection{ Simplifications due to the absence of pressure in shell models }

\label{app_conserved}

In shell models, the dynamical equations for various observables in scale space mimic the Navier-Stokes equations in real space described above,
except that there is no pressure terms. As a consequence, in the absence of external force $f=0$ and in the absence of dissipation by the viscosity $\nu=0$,
the energy $\left( \frac{ \vec v_t^2(\vec r)}{2} \right)  $ and the angular momentum $\vec L_t(\vec r \vert \vec r_0)  $
are conserved quantities satisfying the continuity equations \ref{naviere} and \ref{navierl} respectively.

%%%%%%%%%%%%%%%%%%%%%%%%%%%%%%%%%%%%%%%%%%%%%%%%%%%%%%%%
% REFERENCES

\end{document}